\documentclass[twocolumn,aps,prc,superscriptaddress,showpacs,floatfix,nofootinbib]{revtex4-1}
\usepackage{url}
\usepackage{cancel}
\usepackage[colorlinks,linkcolor=blue,citecolor=blue,filecolor=black,urlcolor=blue]{hyperref}
\usepackage{epsfig,graphics}
\usepackage{graphicx}
\usepackage{dcolumn}
\usepackage{bm}
\usepackage[usenames]{color}
\usepackage{amssymb}
\usepackage{amsmath}
\usepackage{multirow}
\usepackage{float}
\usepackage{harpoon}
\usepackage{MnSymbol}
\usepackage{appendix}
\usepackage{color}
\usepackage{hyperref}
\usepackage{cleveref}

\begin{document}
\title{Disentangling effects of nucleon size and nucleus structure in relativistic heavy-ion collisions}
\author{Hai-Cheng Wang}
\affiliation{School of Physics Science and Engineering, Tongji University, Shanghai 200092, China}
\author{Song-Jie Li}
\affiliation{School of Physics Science and Engineering, Tongji University, Shanghai 200092, China}
\author{Jun Xu}\email[Correspond to\ ]{junxu@tongji.edu.cn}
\affiliation{School of Physics Science and Engineering, Tongji University, Shanghai 200092, China}
\author{Zhong-Zhou Ren}
\affiliation{School of Physics Science and Engineering, Tongji University, Shanghai 200092, China}
\date{\today}
\begin{abstract}
While relativistic heavy-ion collisions become an alternative way of studying nucleus structure, the accurate extraction of nucleus structure could be hampered by the uncertainty of nucleon size, and the latter has attracted people's attention in the past few years. We have compared the impacts of nuclear size and nucleus structure on deformation probes in relativistic heavy-ion collisions based on a multiphase transport (AMPT) model. With increasing nucleon size, the absolute values of the deformation probes are generally reduced due to smeared initial density fluctuations. In heavy systems such as $^{197}$Au+$^{197}$Au collisions, neglecting the nucleon size could underestimate or overestimate significantly the extracted deformation parameter depending on the used deformation probe, while the scaled anisotropic flow and the scaled Pearson correlation coefficient of flow and transverse momentum are good probes of the nucleus deformation rather insensitive to the nucleon size. In small systems such as $^{16}$O+$^{16}$O collisions, the deformation probes are generally more sensitive to the nucleon size than to the nucleus structure, and the transverse momentum fluctuation less sensitive to detailed nucleus structure may serve as a good probe of the nucleon size.
\end{abstract}
\maketitle

Probing nucleus structure with relativistic heavy-ion collisions has been a hot topic in recent years~\cite{Jia:2022ozr}. Typically, collisions by deformed nuclei may lead to larger anisotropies of the overlap region and stronger fluctuations of the overlap area, compared to those by spherical nuclei. The anisotropic flows, the transverse momentum fluctuation, and their correlations in the kinetic freeze-out stage of relativistic heavy-ion collisions, originating from the initial collision geometry, have been shown to be strongly correlated with the deformation parameter of colliding nuclei~\cite{Jia:2021tzt,Jia:2021qyu}. Based on these scaled relations, deformation parameters of $^{96}$Ru~\cite{Zhang:2021kxj}, $^{96}$Zr~\cite{Zhang:2021kxj}, $^{197}$Au~\cite{Giacalone:2021udy}, and $^{238}$U~\cite{STAR:2024wgy} have been successfully extracted, and the shape of $^{129}$Xe nucleus has shown to be a deformed triaxial ellipsoid~\cite{Bally:2021qys,ATLAS:2022dov,ALICE:2024nqd}. People's interest has now turned to collisions involving light nuclei with large deformation and $\alpha$-cluster structures~\cite{YuanyuanWang:2024sgp,Giacalone:2024luz,Giacalone:2024ixe,Prasad:2024ahm,Lu:2025cni}. While the experimental data of $^{16}$O+$^{16}$O at $\sqrt{s_{NN}}=200$ GeV is still under analysis~\cite{Huang:2023viw}, preliminary studies based on, e.g., the AMPT model have been performed~\cite{Zhao:2024feh,Zhang:2024vkh}.

In the original AMPT model~\cite{Lin:2004en} participant nucleons are treated as point particles, while in reality they have a finite size. The charge radius of proton in the rest frame has been measured to be about 0.84 fm~\cite{Hammer:2019uab}, while the strong-interacting radius is expected to be smaller, e.g., the radius is extracted to be about 0.50 fm from the photoproduction of vector mesons~\cite{PhysRevC.81.025203,PhysRevD.104.054015,PhysRevD.94.034042,PhysRevLett.117.052301}. In the initial condition of hydrodynamic models, the transverse profile of a nucleon is generally described by a Gaussian shape with a width of empirically about $w=0.4$ fm (see, e.g., Ref.~\cite{Schenke:2010rr}). Based on hydrodynamic simulations with Bayesian analysis, this value is consistent with that extracted in Ref.~\cite{PhysRevC.94.024907}, while the analyses in Refs.~\cite{Bernhard:2019bmu,PhysRevC.103.054904,PhysRevC.104.054904,PhysRevC.101.024911,PhysRevLett.126.202301,PhysRevC.103.054909,nijs2021predictionspostdictionsrelativisticlead,PARKKILA2022137485} extracted the width parameter of the nucleon as large as $w=0.8-1.1$ fm. It was proposed to measure the nucleon size through the Pearson coefficient characterizing the correlation between the anisotropic flow and the mean transverse momentum~\cite{Giacalone:2021clp}, and this was further highlighted in a paper by the ALICE Collaboration~\cite{ALICE:2021gxt}. A later analysis on the nucleus-nucleus cross section indicated that $w$ should be below 0.7 fm~\cite{Nijs:2022rme}. It has also been found that a similar sub-nucleon effect may impact the collision dynamics in p+p and p+Pb collision systems based on hydrodynamic~\cite{Mantysaari:2017cni,Welsh:2016siu,Schenke:2014zha} and transport~\cite{Zheng:2021jrr,Zhao:2021bef} simulations. A recent review on the sub-nucleon size effect can be found in Ref.~\cite{Giacalone:2022hnz}. Since the finite size of nucleons may affect the initial geometry of the overlap region in relativistic heavy-ion collisions, it is expected to have some impacts on the deformation probes mentioned above. In the present study, we will compare the relative effects of nucleon size and nucleus structure on deformation probes, and investigate the possibility of disentangling the two effects in $^{197}$Au+$^{197}$Au and $^{16}$O+$^{16}$O collisions at $\sqrt{s_{NN}}=200$ GeV based on the AMPT model.

In the string melting version of the AMPT model, hadrons generated by the Heavy-Ion Jet Interacting Generator model~\cite{Wang:1991hta} from collisions of participant nucleons are converted to partons according to the flavor and spin structures of their valence quarks. The momentum spectrum of these initial partons is described by the Lund string fragmentation function
\begin{equation}\label{lund}
f(z) \propto z^{-1} (1-z)^a \exp(-b m_\perp^2/z),
\end{equation}
with $z$ being the light-cone momentum fraction of the produced hadron of transverse mass $m_\perp$ with respect to that of the fragmenting string, and $a$ and $b$ being two parameters. The coordinates of these partons in the transverse plane are originally set as the same as those of the participant nucleons, or in the middle of colliding nucleons in the case of large momentum transfer. In this way, these participant nucleons are treated as point particles, and their finite-size effect has been neglected. In the present study, we resample the transverse coordinates of initial partons according to a Gaussian distribution
\begin{equation}\label{w}
g(x-x_0,y-y_0) \propto \exp\left[ -\frac{(x-x_0)^2+(y-y_0)^2}{2 w^2}\right],
\end{equation}
with $x_0$ and $y_0$ being its original coordinates in the transverse plane, and $w$ being the width parameter. According to the range of $w$ as mentioned in the introduction, we set $w=0$, 0.4, and 0.8 fm, where $w=0$ represents the original case with point participant nucleons, and $w=0.4$ and 0.8 fm correspond to a nucleon root-mean-square raidus of 0.57 and 1.13 fm, respectively. At $\sqrt{s_{NN}}=200$ GeV considered in the present study, the longitudinal coordinates of initial partons are all set to be zero as in the original AMPT model. The evolution of the partonic phase is described by Zhang's Parton Cascade model~\cite{Zhang:1997ej} including two-body elastic collisions with the differential cross section
\begin{equation}\label{xsection}
\frac{d\sigma}{dt} \approx \frac{9\pi \alpha_s^2}{2(t-\mu^2)^2},
\end{equation}
where $t$ is the standard Mandelstam variable for four-momentum transfer. In the present study, we set the strong coupling constant $\alpha_s$ to be 0.33, and the screening mass $\mu$ to be 3.2 fm$^{-1}$, corresponding to a parton scattering cross section of 1.5 mb. After the kinetic freeze-out of these partons, a spatial coalescence model is used to combine valence quarks into hadrons. A relativistic transport model~\cite{Li:1995pra} including various elastic, inelastic, and decay channels is then used to describe the evolution of the hadronic phase until the kinetic freeze-out of all hadrons.

\begin{figure}[ht]
  \centering
  \includegraphics[scale=0.15]{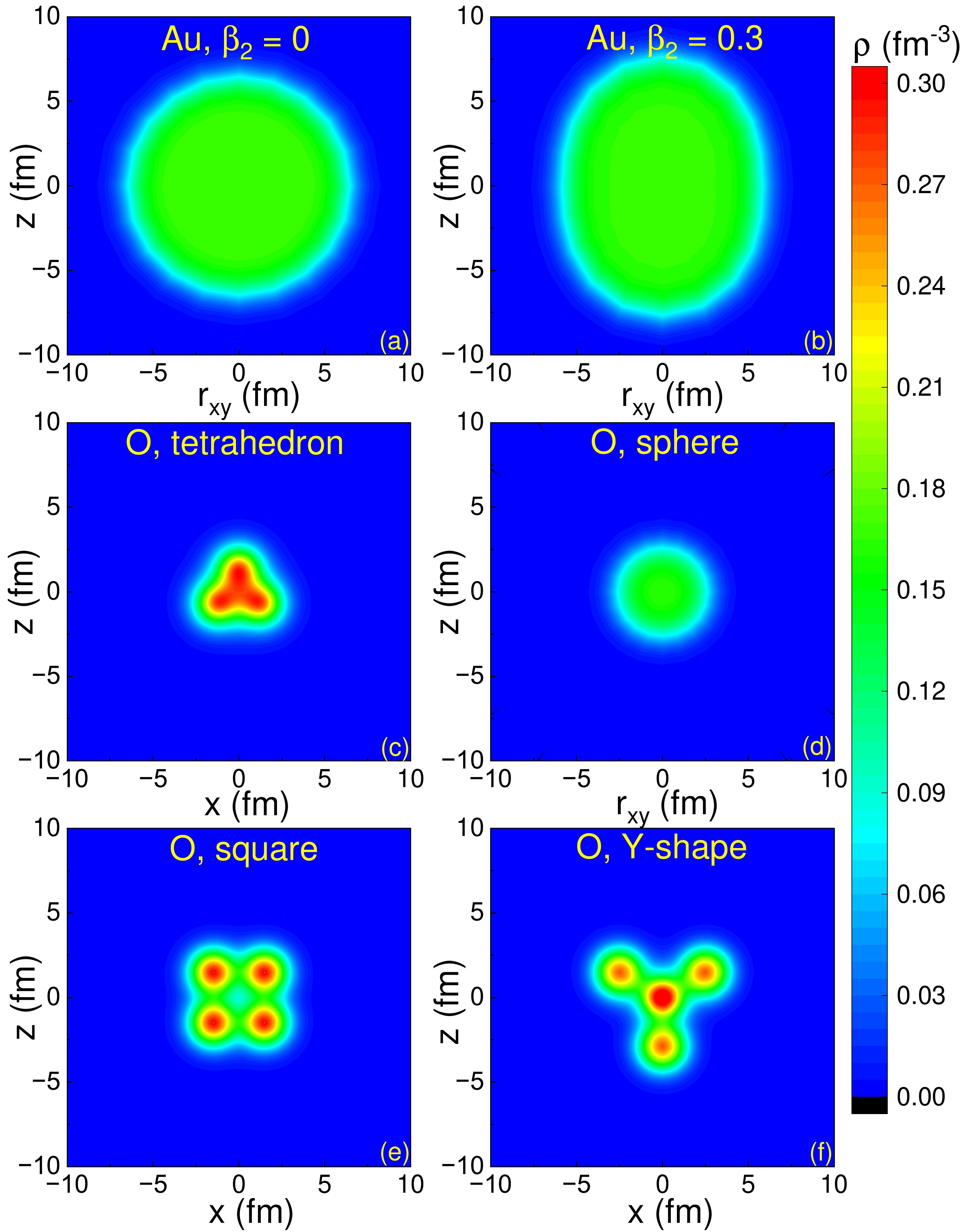}
  \caption{Density contours of $^{197}$Au with $\beta_2=0$ (a) and 0.3 (b) as well as $^{16}$O with tetrahedron (c), sphere (d), square (e), and Y-shape (f) configurations. In panels (a), (b), and (d), $z$ direction represents the symmetric axis, and $r_{xy}$ represents the radius in x-o-y plane.}
  \label{fig0}
\end{figure}

The following Woods-Saxon form is generally used to describe the initial nucleon density distribution of colliding nuclei
\begin{equation}\label{dfms}
\rho(r,\theta) = \frac{\rho_0(1+w_0\frac{r^2}{R_0^2})}{1+\exp\left\{\frac{r-R_0[1+\beta_2 Y_{2,0}(\theta)]}{d}\right\}}.
\end{equation}
In the above, $\rho_0$ is the normalization constant, $R_0$ is the average radius, $d$ is the diffuseness parameter, $w_0$ is the weight parameter, $\beta_2$ is the deformation parameter, and $Y_{2,0}$ is the spherical harmonics. For $^{197}$Au, we use $R_0=6.380$ fm, $d=0.535$ fm, and $w_0=0$, and will compare observables for the spherical ($\beta_2=0$) and deformed ($\beta_2=0.3$) cases. For $^{16}$O, we consider the spherical distribution using $R_0=2.608$ fm, $d=0.513$ fm, $w_0=-0.051$, and $\beta_2=0$, as well as the tetrahedron, square, and Y-shape configurations obtained from the microscopic cluster model with Bloch-Brink wave function~\cite{Liu:2023gun,Wang:2024ulq}. The contours of these density distributions are displayed in Fig.~\ref{fig0}.

\begin{figure}[ht]
  \centering
   \includegraphics[scale=0.35]{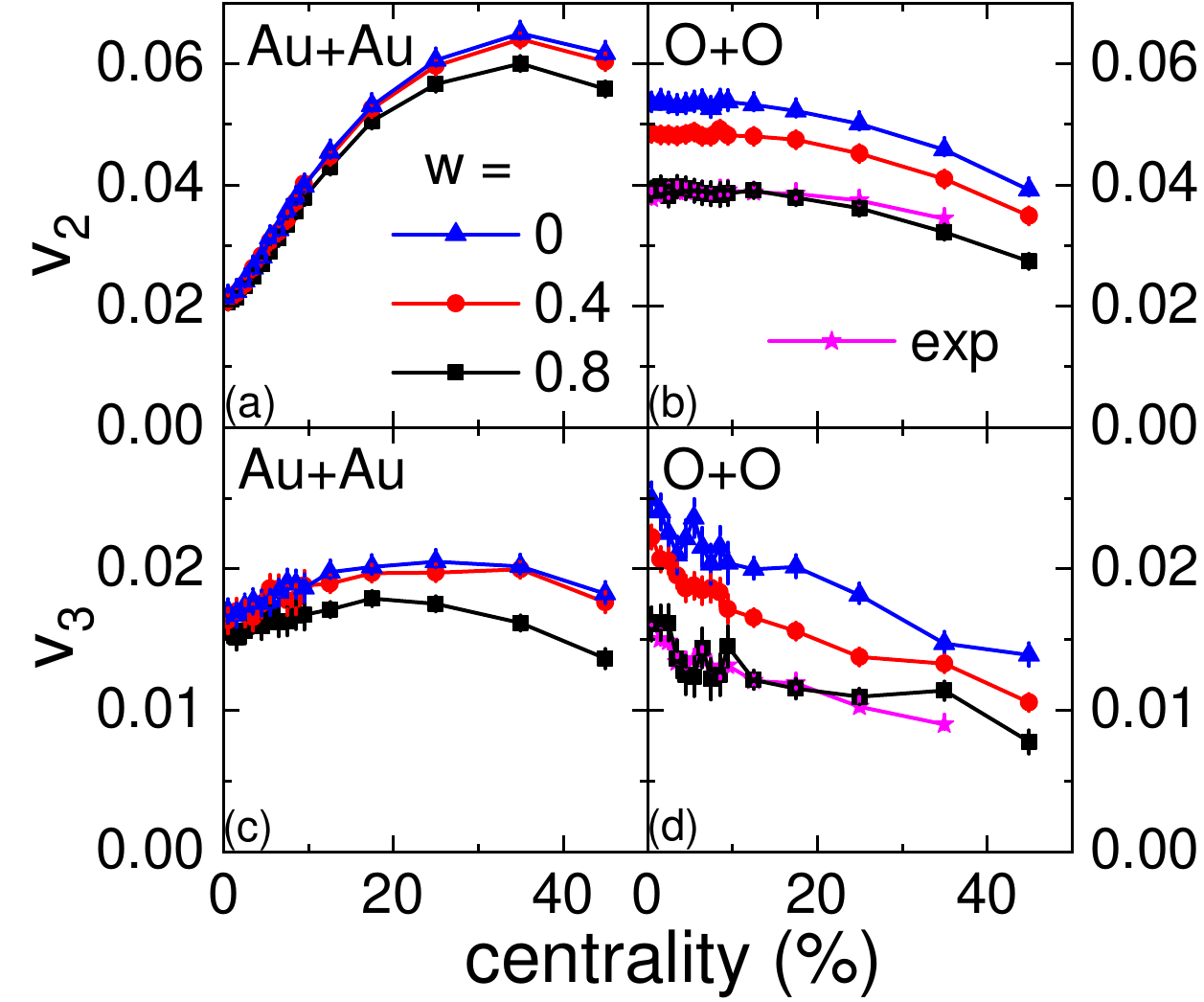}
  \caption{Centrality dependence of the elliptic flow $v_2$ [(a), (b)] and the triangular flow $v_3$ [(c), (d)] in $^{197}$Au+$^{197}$Au collisions [(a), (c)] and $^{16}$O+$^{16}$O collisions [(b), (d)] at $\sqrt{s_{NN}}=200$ GeV. Results with different nucleon sizes characterized by $w$ are compared, and the preliminary experimental data are taken from Refs.~\cite{Huang:2023viw,Zhao:2024feh}.}
  \label{fig1}
\end{figure}

We begin by showing effects of finite nucleon size on anisotropic flows in $^{197}$Au+$^{197}$Au and $^{16}$O+$^{16}$O collisions in Fig.~\ref{fig1}, where the realistic spherical density distribution for $^{197}$Au and energy-favored tetrahedron density distribution for $^{16}$O are used. For $^{197}$Au+$^{197}$Au collisions, we set $a=0.5$ and $b=0.9$ GeV$^{-2}$ in Eq.~(\ref{lund}) as in the previous study~\cite{Xu:2011fe}. For $^{16}$O+$^{16}$O collisions, since the AMPT model has been shown to overestimate the non-flow effect in small collision systems~\cite{PhysRevC.100.024908,PhysRevC.107.024907}, we set $a=2.2$ and $b=0.5$ GeV$^{-2}$ in Eq.~(\ref{lund}), which lead to a soft initial parton momentum spectrum, in order to reduce the non-flow effect. The $n$th-order anisotropic flows $v_n$ are calculated with the same method in the experimental analysis~\cite{Huang:2023viw} from the two-particle cumulant method through
\begin{equation}
v_n\{2\} = \sqrt{c_n} = \sqrt{\langle \cos [n(\varphi_i-\varphi_j)] \rangle_{i,j}},\\
\end{equation}
where $\langle...\rangle_{i,j,...}$ represents the average over all possible combinations of $i$ and $j$ for all events, and $\varphi_i=\arctan(p_{y,i}/p_{x,i})$ represents the momentum direction of the $i$th hadron in the transverse plane. We use information of final charged particles in the range of $|\eta|<1.5$ and $0.2<p_T<2.0$ GeV, and a gap of $|\Delta \eta|>1$ is used in order to suppress the non-flow effect. We have also corrected $v_n$ by subtracting the non-flow effect through~\cite{PhysRevLett.130.242301}
\begin{equation}
c_n \rightarrow c_n - c_n^{peri} \times f,
\end{equation}
where $c_n^{peri}$ represents the contribution from peripheral collisions ($60-80\%$ centralities), and the factor $f$ is taken to be $c_1/c_1^{peri}$. In this way, only $v_n$ in non-peripheral collisions are calculated. Figure~\ref{fig1} shows that all flows are reduced with increasing nucleon size, due to less initial density fluctuations smeared by a larger $w$. The effect is larger in non-central than central $^{197}$Au+$^{197}$Au collisions, and is larger in small systems such as $^{16}$O+$^{16}$O collisions. Instead of assuming a long parton formation time as in Ref.~\cite{Zhao:2024feh}, we found that a finite nucleon size with $w=0.8$ fm reproduces nicely the preliminary results of $v_2$ and $v_3$ measured by the STAR Collaboration~\cite{Huang:2023viw}.

\begin{figure}[ht]
  \centering
  \includegraphics[scale=0.35]{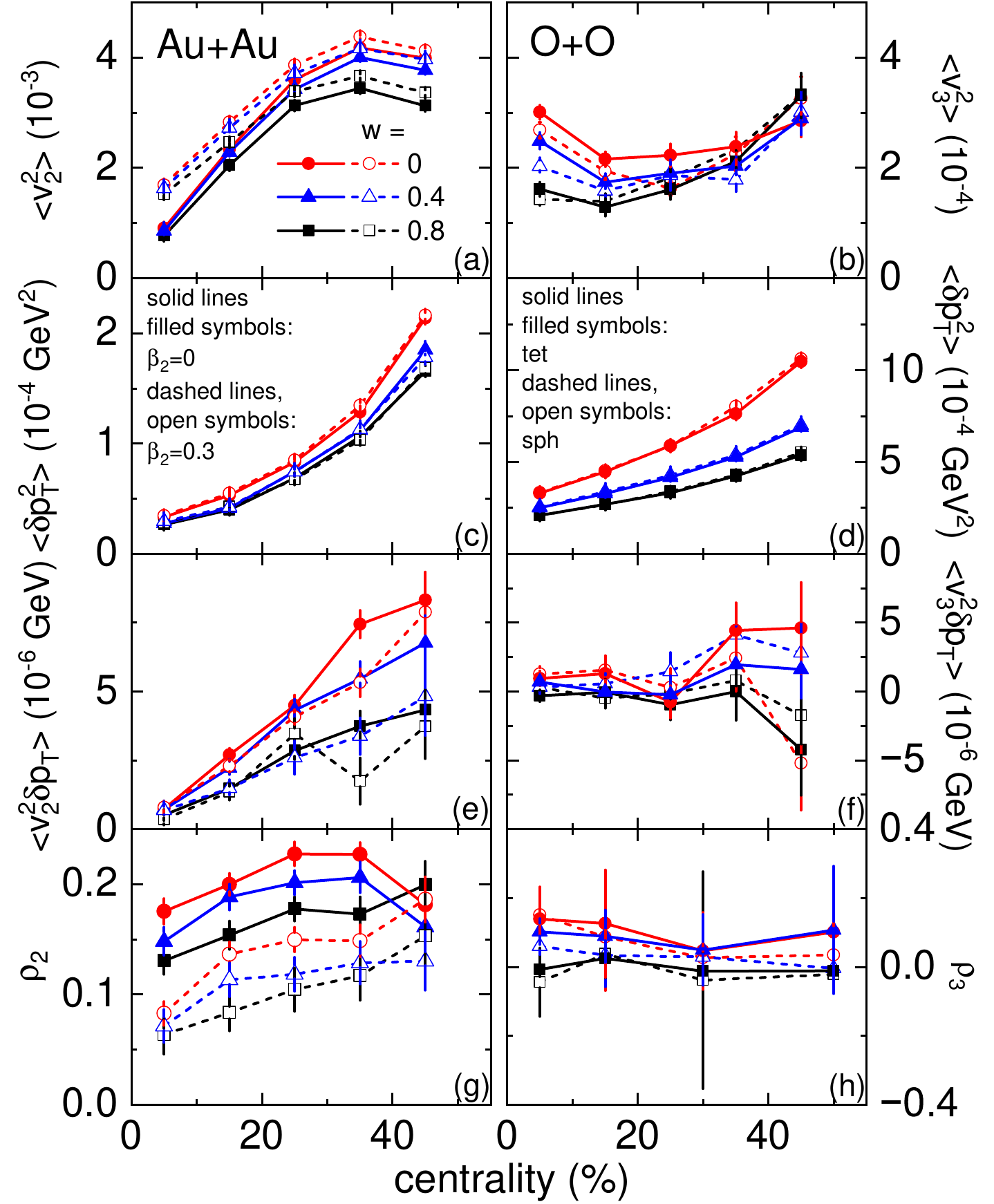}
  \caption{Left: Centrality dependence of $\langle v_2^2 \rangle$ (a), $\langle \delta p_T^2\rangle$ (c), $\langle v_2^2\delta p_T\rangle$ (e), and $\rho_2$ (g) in $^{197}$Au+$^{197}$Au collisions at $\sqrt{s_{NN}}=200$ GeV with different deformation parameters $\beta_2$ and nucleon sizes characterized by $w$; Right: Centrality dependence of $\langle v_3^2 \rangle$ (b), $\langle \delta p_T^2\rangle$ (d), $\langle v_3^2\delta p_T\rangle$ (f), and $\rho_3$ (h) in $^{16}$O+$^{16}$O collisions at $\sqrt{s_{NN}}=200$ GeV with initially sphere (sph) and tetrahedron (tet) configurations and different nucleon sizes characterized by $w$.}
  \label{fig2}
\end{figure}

The effects of nucleus structure and nucleon size on the deformation probes in non-peripheral $^{197}$Au+$^{197}$Au and $^{16}$O+$^{16}$O collisions are extensively compared in Fig.~\ref{fig2}. Here we investigate the following deformation probes calculated according to
\begin{eqnarray}
\langle v_n^2 \rangle &=& \langle \cos [n(\varphi_i-\varphi_j)] \rangle_{i,j}, \\
\langle \delta p_T^2 \rangle &=& \langle (p_{T,i} - \langle \overline{p_T}\rangle) (p_{T,j} - \langle \overline{p_T}\rangle) \rangle_{i,j}, \\
\langle v_n^2 \delta p_T \rangle &=& \langle \cos[n(\varphi_i-\varphi_j)] (p_{T,k} - \langle \overline{p_T}\rangle)\rangle_{i,j,k}, \\
\rho_n &=& \frac{\langle v_n^2 \delta p_T \rangle}{\sqrt{\langle (\delta v_n^2)^2\rangle \langle \delta p_T^2 \rangle}}, \label{rho2}
\end{eqnarray}
respectively. In the above, $p_{T,i}=\sqrt{p_{x,i}^2+p_{y,i}^2}$ is the momentum of the $i$th particle in the transverse plane, $\overline{p_T}$ represents the mean transverse momentum in one event, and $\langle \overline{p_T}\rangle$ represents the averaged value of $\overline{p_T}$ for all events. Four particle cumulant method is used in evaluating $\langle (\delta v_n^2)^2\rangle =\langle v_n^4 \rangle - \langle v_n^2\rangle^2$~\cite{Bozek:2016yoj}. A larger range of $0.2<p_T<3$ GeV for final hadrons is used in order to have a better statistics, particles within $-0.5<\eta<0.5$ are used to evaluate the transverse momentum fluctuation, and particles selected respectively from $1<\eta<2$ and $-2<\eta<-1$ are used to evaluate the anisotropic flow. Since we assume a quadrupole deformation for $^{197}$Au, we compare the deformation probes such as $\langle v_2^2 \rangle$, $\langle \delta p_T^2\rangle$, $\langle v_2^2\delta p_T\rangle$, and $\rho_2$ with different nucleon sizes characterized by $w$. It is seen that a larger $\beta_2$ increases $\langle v_2^2 \rangle$ and $\langle \delta p_T^2\rangle$ but reduces $\langle v_2^2\delta p_T\rangle$ and $\rho_2$, while a larger nucleon size reduces these quantities due to smeared initial density fluctuations. Compared to the effect of nucleus deformation, the effect of $w$ on the absolute value of these quantities is larger for $\langle \delta p_T^2\rangle$ and comparable for $\langle v_2^2 \rangle$ and $\langle v_2^2\delta p_T\rangle$. Such effect is larger at larger centralities, except $\rho_2$ which is basically reduces at all centralities with increasing $w$. Since the tetrahedron configuration of $^{16}$O has a octupole-like deformation, we compare the deformation probes such as $\langle v_3^2 \rangle$, $\langle \delta p_T^2\rangle$, $\langle v_3^2\delta p_T\rangle$, and $\rho_3$ with different nucleon sizes characterized by $w$. Compared to the sphere configuration, it is seen that $\langle v_3^2 \rangle$ is larger for the tetrahedron configuration especially in central collisions, while for other quantities different configurations lead to similar results within statistical errors. On the other hand, the nucleon size has generally larger effects than the nucleus structure on these observables in $^{16}$O+$^{16}$O collisions.

\begin{figure}[ht]
  \centering
  \includegraphics[scale=0.17]{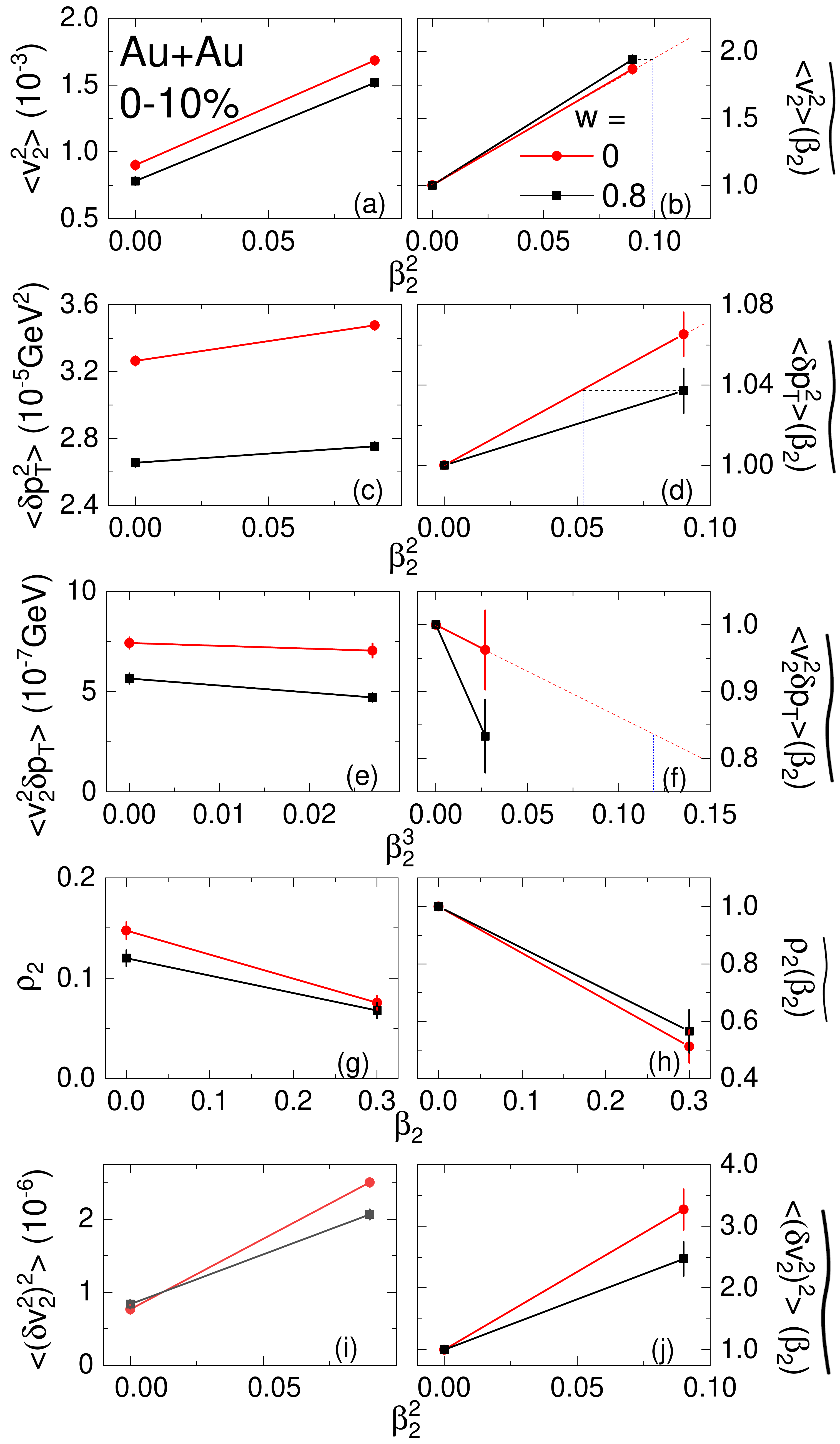}
  \caption{Left: $\beta_2$ dependence of $\langle v_2^2 \rangle$ (a), $\langle \delta p_T^2\rangle$ (c), $\langle v_2^2\delta p_T\rangle$ (e), $\rho_2$ (g), and $\langle (\delta v_2^2)^2 \rangle$ (i) in central ($0-10\%$) $^{197}$Au+$^{197}$Au collisions at $\sqrt{s_{NN}}=200$ GeV with different nucleon sizes characterized by $w$; Right: Similar to the left cases but scaled by the corresponding quantities for $\beta_2=0$, with the dashed lines denoting how possible errors will appear in extracting $\beta_2$ (see text for details).}
  \label{fig3}
\end{figure}

In central collisions of heavy nuclei at relativistic energies, the observables shown in Fig.~\ref{fig2} have the following relations with the deformation parameter~\cite{Jia:2021tzt,Jia:2021qyu,PhysRevC.105.014906}
\begin{eqnarray}
\langle v_2^2 \rangle (\beta_2) &=& \langle v_2^2 \rangle (0) + k_1 \beta_2^2, \label{v2s} \\
\langle \delta p_T^2 \rangle (\beta_2) &=& \langle \delta p_T^2\rangle (0) + k_2 \beta_2^2, \label{pts} \\
\langle v_2^2\delta p_T\rangle (\beta_2) &=& \langle v_2^2\delta p_T\rangle (0) + k_3 \beta_2^3, \label{v2pts}\\
\rho_2(\beta_2) &=& \rho_2(0) + k_4 \beta_2, \label{rho2s}
\end{eqnarray}
where $k_{1 \sim 4}$ are constant coefficients. In the left panels of Fig.~\ref{fig3}, the above relations are illustrated by results from $\beta_2=0$ and 0.3 in central $^{197}$Au+$^{197}$Au collisions, and results from different nucleon sizes characterized by $w$ are compared. It is seen that increasing $w$ generally reduces the corresponding values at $\beta_2=0$ and changes the slopes as well. In real analyses~\cite{Zhang:2021kxj,Giacalone:2021udy,STAR:2024wgy}, the scaled deformation probes, i.e., the ratios of the deformation probes for the interested nucleus to those for a given nucleus of similar mass with known deformation, are used. The relations between $\beta_2$ and the scaled deformation probes, which are taken to be the ratios of the deformation probes $A$ for $\beta_2=0.3$ to $\beta_2=0$, i.e, $\widetilde{A(\beta_2)}=A(\beta_2)/A(0)$, are shown in the right panels of Fig.~\ref{fig3} with different $w$. The effect of $w$ on the slope, which is used to extract the deformation parameter $\beta_2$, depends on the used deformation probe. If the result from $w=0.8$ is the realistic case, treating nucleons as point particles means assuming that the scaled probes follow the trend of $w=0$ according to Eqs.~(\ref{v2s}), (\ref{pts}), and (\ref{v2pts}) denoted by the red dashed lines, and this may lead to the extracted $\beta_2$ value to be about 0.315 from $\widetilde{\langle v_2^2\rangle(\beta_2)}$, 0.228 from $\widetilde{\langle \delta p_T^2\rangle(\beta_2)}$, and 0.495 from $\widetilde{\langle v_2^2\delta p_T\rangle(\beta_2)}$, as shown by the black dashed lines and the blue dotted lines. Therefore, if the nucleon has a large size of $w=0.8$ fm, assuming point nucleons and using scaled $\langle v_2^2\rangle(\beta_2)$, $\langle \delta p_T^2\rangle(\beta_2)$, and $\langle v_2^2\delta p_T\rangle(\beta_2)$ may overestimate $\beta_2$ by $5\%$, underestimate $\beta_2$ by $24\%$, and overestimate $\beta_2$ by $65\%$, respectively, compared to the real value of $\beta_2=0.3$. Therefore, the anisotropic flow looks like a relatively good probe, since the increasing nucleon size reduces the anisotropic flow for different $\beta_2$ in a simultaneous way. On the other hand, the scaled probes involving the transverse momentum fluctuation are more sensitive to initial density fluctuations affected strongly by the finite nucleon size. For $\widetilde{\rho_2(\beta_2)}$, it is more sensitive to the nucleus deformation but insensitive to $w$ within statistical error. To understand this, we have further investigated the behavior of $\langle (\delta v_2^2)^2 \rangle$ in the bottom row of Fig.~\ref{fig3}. As can be seen from Eq.~(\ref{rho2}), the effect of $w$ on the scaled $\rho_2$ is seen to be largely cancelled by the combined effect of $w$ on the scaled $\langle v_2^2 \delta p_T \rangle$, $\langle \delta p_T^2 \rangle$, and $\langle (\delta v_2^2)^2\rangle$.

\begin{figure}[ht]
  \centering
  \includegraphics[scale=0.35]{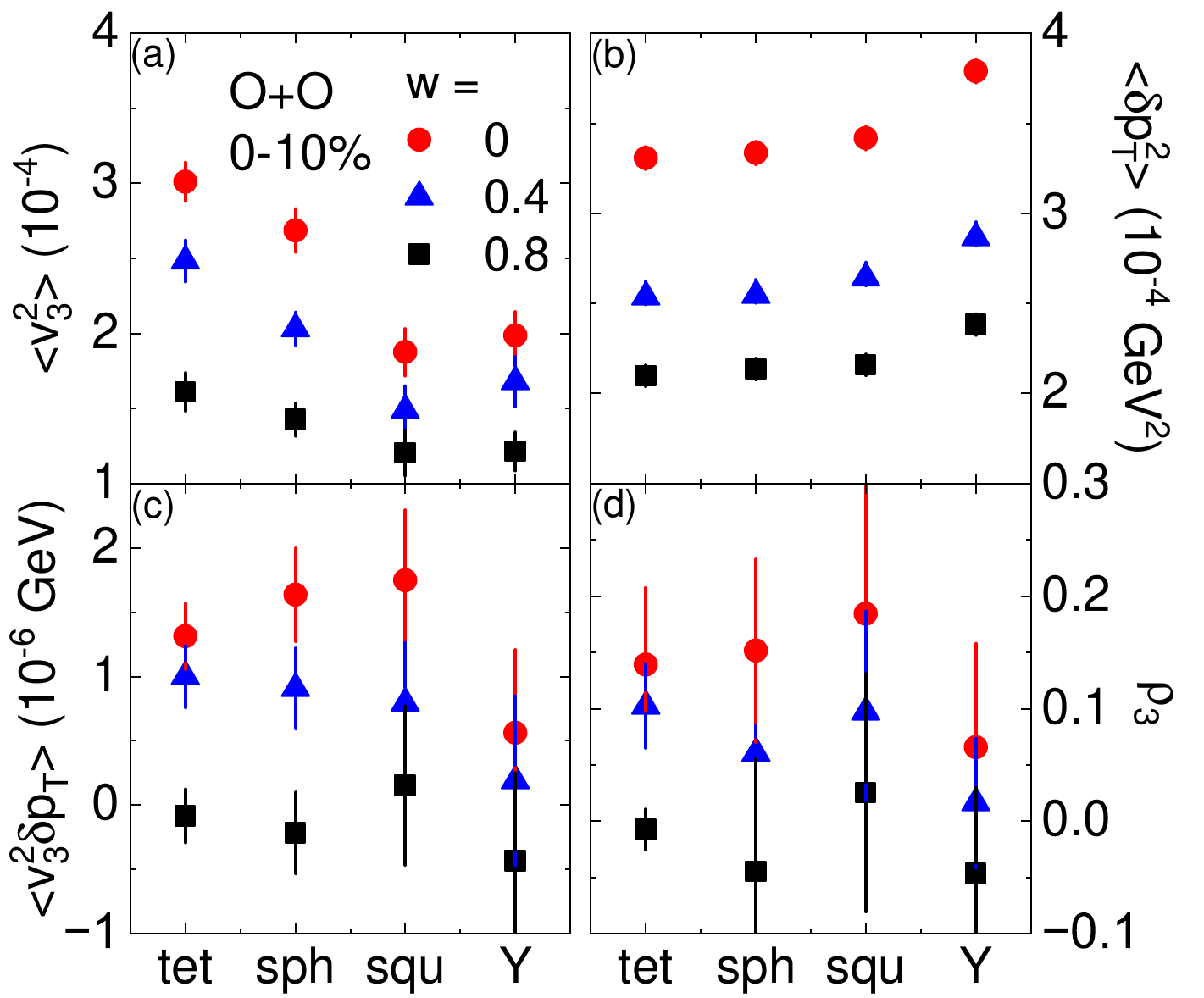}
  \caption{$\langle v_3^2 \rangle$ (a), $\langle \delta p_T^2\rangle$ (b), $\langle v_3^2\delta p_T\rangle$ (c), and $\rho_3$ (d) in $^{16}$O+$^{16}$O collisions for initially sphere (sph), tetrahedron (tet), square (squ), and Y-shape (Y) configurations at $\sqrt{s_{NN}}=200$ GeV with different nucleon sizes characterized by $w$.}
  \label{fig4}
\end{figure}

The deformation probes in central $^{16}$O+$^{16}$O collisions with different initial configurations are shown in Fig.~\ref{fig4}, where results from different nucleon sizes characterized by $w$ are compared. Increasing $w$ generally reduces the values of these quantities due to smeared initial density fluctuations. For all values of $w$, the tetrahedron configuration leads to the largest $\langle v_3^2 \rangle$, and the Y-shape configuration leads to the largest $\langle \delta p_T^2\rangle$, compared with other configurations. For $\langle v_3^2\delta p_T\rangle$ and $\rho_3$, the relative difference among results from different configurations depends on $w$. Overall, the nucleon size has a larger impact on the final observables than the initial nucleus structure in central $^{16}$O+$^{16}$O collisions. For the transverse momentum fluctuation $\langle \delta p_T^2\rangle$, different configurations lead to smaller difference (see also Ref.~\cite{Zhang:2024vkh}) than the effect due to $w$, making it a possible probe of the nucleon size in small collision systems. For other deformation probes such as $\langle v_2^2 \rangle$, $\langle v_2^2\delta p_T\rangle$, and $\rho_2$ not shown here, we found that they are also relatively sensitive to the structure of $^{16}$O, and are not robust probes of $w$ as $\langle \delta p_T^2\rangle$.


To summarize, we have compared the effects of the finite nucleon size and the initial nucleus structure on final deformation probes in $^{197}$Au+$^{197}$Au and $^{16}$O+$^{16}$O collisions at top RHIC energy based on the AMPT model. The preliminary flow data of $^{16}$O+$^{16}$O collisions can be reproduced by a large nucleon size. A larger nucleon size generally reduces the absolute values of deformation probes due to smeared initial density fluctuations. In large collision systems such as $^{197}$Au+$^{197}$Au collisions, neglecting the nucleon size may affect significantly the extracted deformation parameter, and taking the ratio of the observables with the similar mass, as in Refs.~\cite{Zhang:2021kxj,Giacalone:2021udy,STAR:2024wgy,Giacalone:2024luz,Mantysaari:2024uwn,STAR:2021mii,ALICE:2024nqd}, may not completely remove the nucleon size effect. If the nucleon root-mean-square radius is as large as 1.13 fm, for collisions of heavy nuclei with a real quadrupole deformation of $\beta_2=0.3$, the extracted $\beta_2$ by assuming point nucleons will be overestimated by $5\%$, underestimated by $24\%$, and overestimated by $65\%$, from the scaled $\langle v_2^2 \rangle$, $\langle \delta p_T^2\rangle$, and $\langle v_2^2\delta p_T\rangle$, respectively. We found that the scaled anisotropic flow and the scaled Pearson correlation coefficient of flow and transverse momentum serve as good probes of the deformation parameter insensitive to the nucleon size. In small collision systems such as $^{16}$O+$^{16}$O collisions, the impact of the nucleon size on final observables is generally larger than the difference due to different possible configurations, and the transverse momentum fluctuation $\langle \delta p_T^2\rangle$, which is relatively less sensitive to the structure of $^{16}$O, could be a useful probe of the nucleon size. A Bayesian analysis by taking into account both uncertainties of nuclear structure and nucleon size is needed when confronting with the experimental data of these deformation probes. Our study is useful for optimizing the investigation on nuclear structure with collisions by heavy nuclei, and provides references for the on-going experimental analysis on collisions with light nuclei.

This work is supported by the National Key Research and Development Program of China under Grant No. 2022YFA1602404, the Strategic Priority Research Program of the Chinese Academy of Sciences under Grant No. XDB34030000, the National Natural Science Foundation of China under Grant Nos. 12375125, 12035011, and 11975167, and the Fundamental Research Funds for the Central Universities.

\bibliography{finitesize}

\begin{thebibliography}{55}%
\makeatletter
\providecommand \@ifxundefined [1]{%
 \@ifx{#1\undefined}
}%
\providecommand \@ifnum [1]{%
 \ifnum #1\expandafter \@firstoftwo
 \else \expandafter \@secondoftwo
 \fi
}%
\providecommand \@ifx [1]{%
 \ifx #1\expandafter \@firstoftwo
 \else \expandafter \@secondoftwo
 \fi
}%
\providecommand \natexlab [1]{#1}%
\providecommand \enquote  [1]{``#1''}%
\providecommand \bibnamefont  [1]{#1}%
\providecommand \bibfnamefont [1]{#1}%
\providecommand \citenamefont [1]{#1}%
\providecommand \href@noop [0]{\@secondoftwo}%
\providecommand \href [0]{\begingroup \@sanitize@url \@href}%
\providecommand \@href[1]{\@@startlink{#1}\@@href}%
\providecommand \@@href[1]{\endgroup#1\@@endlink}%
\providecommand \@sanitize@url [0]{\catcode `\\12\catcode `\$12\catcode
  `\&12\catcode `\#12\catcode `\^12\catcode `\_12\catcode `\%12\relax}%
\providecommand \@@startlink[1]{}%
\providecommand \@@endlink[0]{}%
\providecommand \url  [0]{\begingroup\@sanitize@url \@url }%
\providecommand \@url [1]{\endgroup\@href {#1}{\urlprefix }}%
\providecommand \urlprefix  [0]{URL }%
\providecommand \Eprint [0]{\href }%
\providecommand \doibase [0]{http://dx.doi.org/}%
\providecommand \selectlanguage [0]{\@gobble}%
\providecommand \bibinfo  [0]{\@secondoftwo}%
\providecommand \bibfield  [0]{\@secondoftwo}%
\providecommand \translation [1]{[#1]}%
\providecommand \BibitemOpen [0]{}%
\providecommand \bibitemStop [0]{}%
\providecommand \bibitemNoStop [0]{.\EOS\space}%
\providecommand \EOS [0]{\spacefactor3000\relax}%
\providecommand \BibitemShut  [1]{\csname bibitem#1\endcsname}%
\let\auto@bib@innerbib\@empty
\bibitem [{\citenamefont {Jia}\ \emph {et~al.}(2024)\citenamefont {Jia} \emph
  {et~al.}}]{Jia:2022ozr}%
  \BibitemOpen
  \bibfield  {author} {\bibinfo {author} {\bibfnamefont {J.}~\bibnamefont
  {Jia}} \emph {et~al.},\ }\href {\doibase 10.1007/s41365-024-01589-w}
  {\bibfield  {journal} {\bibinfo  {journal} {Nucl. Sci. Tech.}\ }\textbf
  {\bibinfo {volume} {35}},\ \bibinfo {pages} {220} (\bibinfo {year} {2024})},\
  \Eprint {http://arxiv.org/abs/2209.11042} {arXiv:2209.11042 [nucl-ex]}
  \BibitemShut {NoStop}%
\bibitem [{\citenamefont {Jia}(2022{\natexlab{a}})}]{Jia:2021tzt}%
  \BibitemOpen
  \bibfield  {author} {\bibinfo {author} {\bibfnamefont {J.}~\bibnamefont
  {Jia}},\ }\href {\doibase 10.1103/PhysRevC.105.014905} {\bibfield  {journal}
  {\bibinfo  {journal} {Phys. Rev. C}\ }\textbf {\bibinfo {volume} {105}},\
  \bibinfo {pages} {014905} (\bibinfo {year} {2022}{\natexlab{a}})},\ \Eprint
  {http://arxiv.org/abs/2106.08768} {arXiv:2106.08768 [nucl-th]} \BibitemShut
  {NoStop}%
\bibitem [{\citenamefont {Jia}(2022{\natexlab{b}})}]{Jia:2021qyu}%
  \BibitemOpen
  \bibfield  {author} {\bibinfo {author} {\bibfnamefont {J.}~\bibnamefont
  {Jia}},\ }\href {\doibase 10.1103/PhysRevC.105.044905} {\bibfield  {journal}
  {\bibinfo  {journal} {Phys. Rev. C}\ }\textbf {\bibinfo {volume} {105}},\
  \bibinfo {pages} {044905} (\bibinfo {year} {2022}{\natexlab{b}})},\ \Eprint
  {http://arxiv.org/abs/2109.00604} {arXiv:2109.00604 [nucl-th]} \BibitemShut
  {NoStop}%
\bibitem [{\citenamefont {Zhang}\ and\ \citenamefont
  {Jia}(2022)}]{Zhang:2021kxj}%
  \BibitemOpen
  \bibfield  {author} {\bibinfo {author} {\bibfnamefont {C.}~\bibnamefont
  {Zhang}}\ and\ \bibinfo {author} {\bibfnamefont {J.}~\bibnamefont {Jia}},\
  }\href {\doibase 10.1103/PhysRevLett.128.022301} {\bibfield  {journal}
  {\bibinfo  {journal} {Phys. Rev. Lett.}\ }\textbf {\bibinfo {volume} {128}},\
  \bibinfo {pages} {022301} (\bibinfo {year} {2022})},\ \Eprint
  {http://arxiv.org/abs/2109.01631} {arXiv:2109.01631 [nucl-th]} \BibitemShut
  {NoStop}%
\bibitem [{\citenamefont {Giacalone}\ \emph {et~al.}(2021)\citenamefont
  {Giacalone}, \citenamefont {Jia},\ and\ \citenamefont
  {Zhang}}]{Giacalone:2021udy}%
  \BibitemOpen
  \bibfield  {author} {\bibinfo {author} {\bibfnamefont {G.}~\bibnamefont
  {Giacalone}}, \bibinfo {author} {\bibfnamefont {J.}~\bibnamefont {Jia}}, \
  and\ \bibinfo {author} {\bibfnamefont {C.}~\bibnamefont {Zhang}},\ }\href
  {\doibase 10.1103/PhysRevLett.127.242301} {\bibfield  {journal} {\bibinfo
  {journal} {Phys. Rev. Lett.}\ }\textbf {\bibinfo {volume} {127}},\ \bibinfo
  {pages} {242301} (\bibinfo {year} {2021})},\ \Eprint
  {http://arxiv.org/abs/2105.01638} {arXiv:2105.01638 [nucl-th]} \BibitemShut
  {NoStop}%
\bibitem [{\citenamefont {Abdulhamid}\ \emph {et~al.}(2024)\citenamefont
  {Abdulhamid} \emph {et~al.}}]{STAR:2024wgy}%
  \BibitemOpen
  \bibfield  {author} {\bibinfo {author} {\bibfnamefont {M.~I.}\ \bibnamefont
  {Abdulhamid}} \emph {et~al.} (\bibinfo {collaboration} {STAR}),\ }\href
  {\doibase 10.1038/s41586-024-08097-2} {\bibfield  {journal} {\bibinfo
  {journal} {Nature}\ }\textbf {\bibinfo {volume} {635}},\ \bibinfo {pages}
  {67} (\bibinfo {year} {2024})},\ \Eprint {http://arxiv.org/abs/2401.06625}
  {arXiv:2401.06625 [nucl-ex]} \BibitemShut {NoStop}%
\bibitem [{\citenamefont {Bally}\ \emph {et~al.}(2022)\citenamefont {Bally},
  \citenamefont {Bender}, \citenamefont {Giacalone},\ and\ \citenamefont
  {Som\`a}}]{Bally:2021qys}%
  \BibitemOpen
  \bibfield  {author} {\bibinfo {author} {\bibfnamefont {B.}~\bibnamefont
  {Bally}}, \bibinfo {author} {\bibfnamefont {M.}~\bibnamefont {Bender}},
  \bibinfo {author} {\bibfnamefont {G.}~\bibnamefont {Giacalone}}, \ and\
  \bibinfo {author} {\bibfnamefont {V.}~\bibnamefont {Som\`a}},\ }\href
  {\doibase 10.1103/PhysRevLett.128.082301} {\bibfield  {journal} {\bibinfo
  {journal} {Phys. Rev. Lett.}\ }\textbf {\bibinfo {volume} {128}},\ \bibinfo
  {pages} {082301} (\bibinfo {year} {2022})},\ \Eprint
  {http://arxiv.org/abs/2108.09578} {arXiv:2108.09578 [nucl-th]} \BibitemShut
  {NoStop}%
\bibitem [{\citenamefont {Aad}\ \emph {et~al.}(2023)\citenamefont {Aad} \emph
  {et~al.}}]{ATLAS:2022dov}%
  \BibitemOpen
  \bibfield  {author} {\bibinfo {author} {\bibfnamefont {G.}~\bibnamefont
  {Aad}} \emph {et~al.} (\bibinfo {collaboration} {ATLAS}),\ }\href {\doibase
  10.1103/PhysRevC.107.054910} {\bibfield  {journal} {\bibinfo  {journal}
  {Phys. Rev. C}\ }\textbf {\bibinfo {volume} {107}},\ \bibinfo {pages}
  {054910} (\bibinfo {year} {2023})},\ \Eprint
  {http://arxiv.org/abs/2205.00039} {arXiv:2205.00039 [nucl-ex]} \BibitemShut
  {NoStop}%
\bibitem [{\citenamefont {Acharya}\ \emph {et~al.}(2024)\citenamefont {Acharya}
  \emph {et~al.}}]{ALICE:2024nqd}%
  \BibitemOpen
  \bibfield  {author} {\bibinfo {author} {\bibfnamefont {S.}~\bibnamefont
  {Acharya}} \emph {et~al.} (\bibinfo {collaboration} {ALICE}),\ }\href@noop {}
  {\  (\bibinfo {year} {2024})},\ \Eprint {http://arxiv.org/abs/2409.04343}
  {arXiv:2409.04343 [nucl-ex]} \BibitemShut {NoStop}%
\bibitem [{\citenamefont {Wang}\ \emph
  {et~al.}(2024{\natexlab{a}})\citenamefont {Wang}, \citenamefont {Zhao},
  \citenamefont {Cao}, \citenamefont {Xu},\ and\ \citenamefont
  {Song}}]{YuanyuanWang:2024sgp}%
  \BibitemOpen
  \bibfield  {author} {\bibinfo {author} {\bibfnamefont {Y.}~\bibnamefont
  {Wang}}, \bibinfo {author} {\bibfnamefont {S.}~\bibnamefont {Zhao}}, \bibinfo
  {author} {\bibfnamefont {B.}~\bibnamefont {Cao}}, \bibinfo {author}
  {\bibfnamefont {H.-j.}\ \bibnamefont {Xu}}, \ and\ \bibinfo {author}
  {\bibfnamefont {H.}~\bibnamefont {Song}},\ }\href {\doibase
  10.1103/PhysRevC.109.L051904} {\bibfield  {journal} {\bibinfo  {journal}
  {Phys. Rev. C}\ }\textbf {\bibinfo {volume} {109}},\ \bibinfo {pages}
  {L051904} (\bibinfo {year} {2024}{\natexlab{a}})},\ \Eprint
  {http://arxiv.org/abs/2401.15723} {arXiv:2401.15723 [nucl-th]} \BibitemShut
  {NoStop}%
\bibitem [{\citenamefont {Giacalone}\ \emph {et~al.}()\citenamefont {Giacalone}
  \emph {et~al.}}]{Giacalone:2024luz}%
  \BibitemOpen
  \bibfield  {author} {\bibinfo {author} {\bibfnamefont {G.}~\bibnamefont
  {Giacalone}} \emph {et~al.},\ }\href@noop {} {\ }\Eprint
  {http://arxiv.org/abs/2402.05995} {arXiv:2402.05995 [nucl-th]} \BibitemShut
  {NoStop}%
\bibitem [{\citenamefont {Giacalone}\ \emph {et~al.}(2025)\citenamefont
  {Giacalone} \emph {et~al.}}]{Giacalone:2024ixe}%
  \BibitemOpen
  \bibfield  {author} {\bibinfo {author} {\bibfnamefont {G.}~\bibnamefont
  {Giacalone}} \emph {et~al.},\ }\href {\doibase
  10.1103/PhysRevLett.134.082301} {\bibfield  {journal} {\bibinfo  {journal}
  {Phys. Rev. Lett.}\ }\textbf {\bibinfo {volume} {134}},\ \bibinfo {pages}
  {082301} (\bibinfo {year} {2025})},\ \Eprint
  {http://arxiv.org/abs/2405.20210} {arXiv:2405.20210 [nucl-th]} \BibitemShut
  {NoStop}%
\bibitem [{\citenamefont {Prasad}\ \emph {et~al.}(2025)\citenamefont {Prasad},
  \citenamefont {Mallick}, \citenamefont {Sahoo},\ and\ \citenamefont
  {Barnaf\"oldi}}]{Prasad:2024ahm}%
  \BibitemOpen
  \bibfield  {author} {\bibinfo {author} {\bibfnamefont {S.}~\bibnamefont
  {Prasad}}, \bibinfo {author} {\bibfnamefont {N.}~\bibnamefont {Mallick}},
  \bibinfo {author} {\bibfnamefont {R.}~\bibnamefont {Sahoo}}, \ and\ \bibinfo
  {author} {\bibfnamefont {G.~G.}\ \bibnamefont {Barnaf\"oldi}},\ }\href
  {\doibase 10.1016/j.physletb.2024.139145} {\bibfield  {journal} {\bibinfo
  {journal} {Phys. Lett. B}\ }\textbf {\bibinfo {volume} {860}},\ \bibinfo
  {pages} {139145} (\bibinfo {year} {2025})},\ \Eprint
  {http://arxiv.org/abs/2407.15065} {arXiv:2407.15065 [nucl-th]} \BibitemShut
  {NoStop}%
\bibitem [{\citenamefont {Lu}\ \emph {et~al.}()\citenamefont {Lu},
  \citenamefont {Zhao}, \citenamefont {Nielsen}, \citenamefont {Li},\ and\
  \citenamefont {Zhou}}]{Lu:2025cni}%
  \BibitemOpen
  \bibfield  {author} {\bibinfo {author} {\bibfnamefont {Z.}~\bibnamefont
  {Lu}}, \bibinfo {author} {\bibfnamefont {M.}~\bibnamefont {Zhao}}, \bibinfo
  {author} {\bibfnamefont {E.~G.~D.}\ \bibnamefont {Nielsen}}, \bibinfo
  {author} {\bibfnamefont {X.}~\bibnamefont {Li}}, \ and\ \bibinfo {author}
  {\bibfnamefont {Y.}~\bibnamefont {Zhou}},\ }\href@noop {} {\ }\Eprint
  {http://arxiv.org/abs/2501.14852} {arXiv:2501.14852 [nucl-th]} \BibitemShut
  {NoStop}%
\bibitem [{\citenamefont {Huang}()}]{Huang:2023viw}%
  \BibitemOpen
  \bibfield  {author} {\bibinfo {author} {\bibfnamefont {S.}~\bibnamefont
  {Huang}},\ }\Eprint {http://arxiv.org/abs/2312.12167} {arXiv:2312.12167
  [nucl-ex]} \BibitemShut {NoStop}%
\bibitem [{\citenamefont {Zhao}\ \emph {et~al.}()\citenamefont {Zhao},
  \citenamefont {Ma}, \citenamefont {Zhou}, \citenamefont {Lin},\ and\
  \citenamefont {Zhang}}]{Zhao:2024feh}%
  \BibitemOpen
  \bibfield  {author} {\bibinfo {author} {\bibfnamefont {X.-L.}\ \bibnamefont
  {Zhao}}, \bibinfo {author} {\bibfnamefont {G.-L.}\ \bibnamefont {Ma}},
  \bibinfo {author} {\bibfnamefont {Y.}~\bibnamefont {Zhou}}, \bibinfo {author}
  {\bibfnamefont {Z.-W.}\ \bibnamefont {Lin}}, \ and\ \bibinfo {author}
  {\bibfnamefont {C.}~\bibnamefont {Zhang}},\ }\href@noop {} {\ }\Eprint
  {http://arxiv.org/abs/2404.09780} {arXiv:2404.09780 [nucl-th]} \BibitemShut
  {NoStop}%
\bibitem [{\citenamefont {Zhang}\ \emph {et~al.}()\citenamefont {Zhang},
  \citenamefont {Chen}, \citenamefont {Giacalone}, \citenamefont {Huang},
  \citenamefont {Jia},\ and\ \citenamefont {Ma}}]{Zhang:2024vkh}%
  \BibitemOpen
  \bibfield  {author} {\bibinfo {author} {\bibfnamefont {C.}~\bibnamefont
  {Zhang}}, \bibinfo {author} {\bibfnamefont {J.}~\bibnamefont {Chen}},
  \bibinfo {author} {\bibfnamefont {G.}~\bibnamefont {Giacalone}}, \bibinfo
  {author} {\bibfnamefont {S.}~\bibnamefont {Huang}}, \bibinfo {author}
  {\bibfnamefont {J.}~\bibnamefont {Jia}}, \ and\ \bibinfo {author}
  {\bibfnamefont {Y.-G.}\ \bibnamefont {Ma}},\ }\href@noop {} {\ }\Eprint
  {http://arxiv.org/abs/2404.08385} {arXiv:2404.08385 [nucl-th]} \BibitemShut
  {NoStop}%
\bibitem [{\citenamefont {Lin}\ \emph {et~al.}(2005)\citenamefont {Lin},
  \citenamefont {Ko}, \citenamefont {Li}, \citenamefont {Zhang},\ and\
  \citenamefont {Pal}}]{Lin:2004en}%
  \BibitemOpen
  \bibfield  {author} {\bibinfo {author} {\bibfnamefont {Z.-W.}\ \bibnamefont
  {Lin}}, \bibinfo {author} {\bibfnamefont {C.~M.}\ \bibnamefont {Ko}},
  \bibinfo {author} {\bibfnamefont {B.-A.}\ \bibnamefont {Li}}, \bibinfo
  {author} {\bibfnamefont {B.}~\bibnamefont {Zhang}}, \ and\ \bibinfo {author}
  {\bibfnamefont {S.}~\bibnamefont {Pal}},\ }\href {\doibase
  10.1103/PhysRevC.72.064901} {\bibfield  {journal} {\bibinfo  {journal} {Phys.
  Rev. C}\ }\textbf {\bibinfo {volume} {72}},\ \bibinfo {pages} {064901}
  (\bibinfo {year} {2005})},\ \Eprint {http://arxiv.org/abs/nucl-th/0411110}
  {arXiv:nucl-th/0411110} \BibitemShut {NoStop}%
\bibitem [{\citenamefont {Hammer}\ and\ \citenamefont
  {Mei\ss{}ner}(2020)}]{Hammer:2019uab}%
  \BibitemOpen
  \bibfield  {author} {\bibinfo {author} {\bibfnamefont {H.-W.}\ \bibnamefont
  {Hammer}}\ and\ \bibinfo {author} {\bibfnamefont {U.-G.}\ \bibnamefont
  {Mei\ss{}ner}},\ }\href {\doibase 10.1016/j.scib.2019.12.012} {\bibfield
  {journal} {\bibinfo  {journal} {Sci. Bull.}\ }\textbf {\bibinfo {volume}
  {65}},\ \bibinfo {pages} {257} (\bibinfo {year} {2020})},\ \Eprint
  {http://arxiv.org/abs/1912.03881} {arXiv:1912.03881 [hep-ph]} \BibitemShut
  {NoStop}%
\bibitem [{\citenamefont {Caldwell}\ and\ \citenamefont
  {Kowalski}(2010)}]{PhysRevC.81.025203}%
  \BibitemOpen
  \bibfield  {author} {\bibinfo {author} {\bibfnamefont {A.}~\bibnamefont
  {Caldwell}}\ and\ \bibinfo {author} {\bibfnamefont {H.}~\bibnamefont
  {Kowalski}},\ }\href {\doibase 10.1103/PhysRevC.81.025203} {\bibfield
  {journal} {\bibinfo  {journal} {Phys. Rev. C}\ }\textbf {\bibinfo {volume}
  {81}},\ \bibinfo {pages} {025203} (\bibinfo {year} {2010})}\BibitemShut
  {NoStop}%
\bibitem [{\citenamefont {Kharzeev}(2021)}]{PhysRevD.104.054015}%
  \BibitemOpen
  \bibfield  {author} {\bibinfo {author} {\bibfnamefont {D.~E.}\ \bibnamefont
  {Kharzeev}},\ }\href {\doibase 10.1103/PhysRevD.104.054015} {\bibfield
  {journal} {\bibinfo  {journal} {Phys. Rev. D}\ }\textbf {\bibinfo {volume}
  {104}},\ \bibinfo {pages} {054015} (\bibinfo {year} {2021})}\BibitemShut
  {NoStop}%
\bibitem [{\citenamefont {M\"antysaari}\ and\ \citenamefont
  {Schenke}(2016{\natexlab{a}})}]{PhysRevD.94.034042}%
  \BibitemOpen
  \bibfield  {author} {\bibinfo {author} {\bibfnamefont {H.}~\bibnamefont
  {M\"antysaari}}\ and\ \bibinfo {author} {\bibfnamefont {B.}~\bibnamefont
  {Schenke}},\ }\href {\doibase 10.1103/PhysRevD.94.034042} {\bibfield
  {journal} {\bibinfo  {journal} {Phys. Rev. D}\ }\textbf {\bibinfo {volume}
  {94}},\ \bibinfo {pages} {034042} (\bibinfo {year}
  {2016}{\natexlab{a}})}\BibitemShut {NoStop}%
\bibitem [{\citenamefont {M\"antysaari}\ and\ \citenamefont
  {Schenke}(2016{\natexlab{b}})}]{PhysRevLett.117.052301}%
  \BibitemOpen
  \bibfield  {author} {\bibinfo {author} {\bibfnamefont {H.}~\bibnamefont
  {M\"antysaari}}\ and\ \bibinfo {author} {\bibfnamefont {B.}~\bibnamefont
  {Schenke}},\ }\href {\doibase 10.1103/PhysRevLett.117.052301} {\bibfield
  {journal} {\bibinfo  {journal} {Phys. Rev. Lett.}\ }\textbf {\bibinfo
  {volume} {117}},\ \bibinfo {pages} {052301} (\bibinfo {year}
  {2016}{\natexlab{b}})}\BibitemShut {NoStop}%
\bibitem [{\citenamefont {Schenke}\ \emph {et~al.}(2011)\citenamefont
  {Schenke}, \citenamefont {Jeon},\ and\ \citenamefont
  {Gale}}]{Schenke:2010rr}%
  \BibitemOpen
  \bibfield  {author} {\bibinfo {author} {\bibfnamefont {B.}~\bibnamefont
  {Schenke}}, \bibinfo {author} {\bibfnamefont {S.}~\bibnamefont {Jeon}}, \
  and\ \bibinfo {author} {\bibfnamefont {C.}~\bibnamefont {Gale}},\ }\href
  {\doibase 10.1103/PhysRevLett.106.042301} {\bibfield  {journal} {\bibinfo
  {journal} {Phys. Rev. Lett.}\ }\textbf {\bibinfo {volume} {106}},\ \bibinfo
  {pages} {042301} (\bibinfo {year} {2011})},\ \Eprint
  {http://arxiv.org/abs/1009.3244} {arXiv:1009.3244 [hep-ph]} \BibitemShut
  {NoStop}%
\bibitem [{\citenamefont {Bernhard}\ \emph {et~al.}(2016)\citenamefont
  {Bernhard}, \citenamefont {Moreland}, \citenamefont {Bass}, \citenamefont
  {Liu},\ and\ \citenamefont {Heinz}}]{PhysRevC.94.024907}%
  \BibitemOpen
  \bibfield  {author} {\bibinfo {author} {\bibfnamefont {J.~E.}\ \bibnamefont
  {Bernhard}}, \bibinfo {author} {\bibfnamefont {J.~S.}\ \bibnamefont
  {Moreland}}, \bibinfo {author} {\bibfnamefont {S.~A.}\ \bibnamefont {Bass}},
  \bibinfo {author} {\bibfnamefont {J.}~\bibnamefont {Liu}}, \ and\ \bibinfo
  {author} {\bibfnamefont {U.}~\bibnamefont {Heinz}},\ }\href {\doibase
  10.1103/PhysRevC.94.024907} {\bibfield  {journal} {\bibinfo  {journal} {Phys.
  Rev. C}\ }\textbf {\bibinfo {volume} {94}},\ \bibinfo {pages} {024907}
  (\bibinfo {year} {2016})}\BibitemShut {NoStop}%
\bibitem [{\citenamefont {Bernhard}\ \emph {et~al.}(2019)\citenamefont
  {Bernhard}, \citenamefont {Moreland},\ and\ \citenamefont
  {Bass}}]{Bernhard:2019bmu}%
  \BibitemOpen
  \bibfield  {author} {\bibinfo {author} {\bibfnamefont {J.~E.}\ \bibnamefont
  {Bernhard}}, \bibinfo {author} {\bibfnamefont {J.~S.}\ \bibnamefont
  {Moreland}}, \ and\ \bibinfo {author} {\bibfnamefont {S.~A.}\ \bibnamefont
  {Bass}},\ }\href {\doibase 10.1038/s41567-019-0611-8} {\bibfield  {journal}
  {\bibinfo  {journal} {Nature Phys.}\ }\textbf {\bibinfo {volume} {15}},\
  \bibinfo {pages} {1113} (\bibinfo {year} {2019})}\BibitemShut {NoStop}%
\bibitem [{\citenamefont {Everett}\ \emph {et~al.}(2021)\citenamefont {Everett}
  \emph {et~al.}}]{PhysRevC.103.054904}%
  \BibitemOpen
  \bibfield  {author} {\bibinfo {author} {\bibfnamefont {D.}~\bibnamefont
  {Everett}} \emph {et~al.} (\bibinfo {collaboration} {JETSCAPE
  Collaboration}),\ }\href {\doibase 10.1103/PhysRevC.103.054904} {\bibfield
  {journal} {\bibinfo  {journal} {Phys. Rev. C}\ }\textbf {\bibinfo {volume}
  {103}},\ \bibinfo {pages} {054904} (\bibinfo {year} {2021})}\BibitemShut
  {NoStop}%
\bibitem [{\citenamefont {Parkkila}\ \emph {et~al.}(2021)\citenamefont
  {Parkkila}, \citenamefont {Onnerstad},\ and\ \citenamefont
  {Kim}}]{PhysRevC.104.054904}%
  \BibitemOpen
  \bibfield  {author} {\bibinfo {author} {\bibfnamefont {J.~E.}\ \bibnamefont
  {Parkkila}}, \bibinfo {author} {\bibfnamefont {A.}~\bibnamefont {Onnerstad}},
  \ and\ \bibinfo {author} {\bibfnamefont {D.~J.}\ \bibnamefont {Kim}},\ }\href
  {\doibase 10.1103/PhysRevC.104.054904} {\bibfield  {journal} {\bibinfo
  {journal} {Phys. Rev. C}\ }\textbf {\bibinfo {volume} {104}},\ \bibinfo
  {pages} {054904} (\bibinfo {year} {2021})}\BibitemShut {NoStop}%
\bibitem [{\citenamefont {Moreland}\ \emph {et~al.}(2020)\citenamefont
  {Moreland}, \citenamefont {Bernhard},\ and\ \citenamefont
  {Bass}}]{PhysRevC.101.024911}%
  \BibitemOpen
  \bibfield  {author} {\bibinfo {author} {\bibfnamefont {J.~S.}\ \bibnamefont
  {Moreland}}, \bibinfo {author} {\bibfnamefont {J.~E.}\ \bibnamefont
  {Bernhard}}, \ and\ \bibinfo {author} {\bibfnamefont {S.~A.}\ \bibnamefont
  {Bass}},\ }\href {\doibase 10.1103/PhysRevC.101.024911} {\bibfield  {journal}
  {\bibinfo  {journal} {Phys. Rev. C}\ }\textbf {\bibinfo {volume} {101}},\
  \bibinfo {pages} {024911} (\bibinfo {year} {2020})}\BibitemShut {NoStop}%
\bibitem [{\citenamefont {Nijs}\ \emph
  {et~al.}(2021{\natexlab{a}})\citenamefont {Nijs}, \citenamefont {van~der
  Schee}, \citenamefont {G\"ursoy},\ and\ \citenamefont
  {Snellings}}]{PhysRevLett.126.202301}%
  \BibitemOpen
  \bibfield  {author} {\bibinfo {author} {\bibfnamefont {G.}~\bibnamefont
  {Nijs}}, \bibinfo {author} {\bibfnamefont {W.}~\bibnamefont {van~der Schee}},
  \bibinfo {author} {\bibfnamefont {U.}~\bibnamefont {G\"ursoy}}, \ and\
  \bibinfo {author} {\bibfnamefont {R.}~\bibnamefont {Snellings}},\ }\href
  {\doibase 10.1103/PhysRevLett.126.202301} {\bibfield  {journal} {\bibinfo
  {journal} {Phys. Rev. Lett.}\ }\textbf {\bibinfo {volume} {126}},\ \bibinfo
  {pages} {202301} (\bibinfo {year} {2021}{\natexlab{a}})}\BibitemShut
  {NoStop}%
\bibitem [{\citenamefont {Nijs}\ \emph
  {et~al.}(2021{\natexlab{b}})\citenamefont {Nijs}, \citenamefont {van~der
  Schee}, \citenamefont {G\"ursoy},\ and\ \citenamefont
  {Snellings}}]{PhysRevC.103.054909}%
  \BibitemOpen
  \bibfield  {author} {\bibinfo {author} {\bibfnamefont {G.}~\bibnamefont
  {Nijs}}, \bibinfo {author} {\bibfnamefont {W.}~\bibnamefont {van~der Schee}},
  \bibinfo {author} {\bibfnamefont {U.}~\bibnamefont {G\"ursoy}}, \ and\
  \bibinfo {author} {\bibfnamefont {R.}~\bibnamefont {Snellings}},\ }\href
  {\doibase 10.1103/PhysRevC.103.054909} {\bibfield  {journal} {\bibinfo
  {journal} {Phys. Rev. C}\ }\textbf {\bibinfo {volume} {103}},\ \bibinfo
  {pages} {054909} (\bibinfo {year} {2021}{\natexlab{b}})}\BibitemShut
  {NoStop}%
\bibitem [{\citenamefont {Nijs}\ and\ \citenamefont {van~der
  Schee}()}]{nijs2021predictionspostdictionsrelativisticlead}%
  \BibitemOpen
  \bibfield  {author} {\bibinfo {author} {\bibfnamefont {G.}~\bibnamefont
  {Nijs}}\ and\ \bibinfo {author} {\bibfnamefont {W.}~\bibnamefont {van~der
  Schee}},\ }\href {https://arxiv.org/abs/2110.13153} {}\Eprint
  {http://arxiv.org/abs/2110.13153} {arXiv:2110.13153 [nucl-th]} \BibitemShut
  {NoStop}%
\bibitem [{\citenamefont {Parkkila}\ \emph {et~al.}(2022)\citenamefont
  {Parkkila}, \citenamefont {Onnerstad}, \citenamefont {Taghavi}, \citenamefont
  {Mordasini}, \citenamefont {Bilandzic}, \citenamefont {Virta},\ and\
  \citenamefont {Kim}}]{PARKKILA2022137485}%
  \BibitemOpen
  \bibfield  {author} {\bibinfo {author} {\bibfnamefont {J.}~\bibnamefont
  {Parkkila}}, \bibinfo {author} {\bibfnamefont {A.}~\bibnamefont {Onnerstad}},
  \bibinfo {author} {\bibfnamefont {S.}~\bibnamefont {Taghavi}}, \bibinfo
  {author} {\bibfnamefont {C.}~\bibnamefont {Mordasini}}, \bibinfo {author}
  {\bibfnamefont {A.}~\bibnamefont {Bilandzic}}, \bibinfo {author}
  {\bibfnamefont {M.}~\bibnamefont {Virta}}, \ and\ \bibinfo {author}
  {\bibfnamefont {D.}~\bibnamefont {Kim}},\ }\href {\doibase
  https://doi.org/10.1016/j.physletb.2022.137485} {\bibfield  {journal}
  {\bibinfo  {journal} {Physics Letters B}\ }\textbf {\bibinfo {volume}
  {835}},\ \bibinfo {pages} {137485} (\bibinfo {year} {2022})}\BibitemShut
  {NoStop}%
\bibitem [{\citenamefont {Giacalone}\ \emph {et~al.}(2022)\citenamefont
  {Giacalone}, \citenamefont {Schenke},\ and\ \citenamefont
  {Shen}}]{Giacalone:2021clp}%
  \BibitemOpen
  \bibfield  {author} {\bibinfo {author} {\bibfnamefont {G.}~\bibnamefont
  {Giacalone}}, \bibinfo {author} {\bibfnamefont {B.}~\bibnamefont {Schenke}},
  \ and\ \bibinfo {author} {\bibfnamefont {C.}~\bibnamefont {Shen}},\ }\href
  {\doibase 10.1103/PhysRevLett.128.042301} {\bibfield  {journal} {\bibinfo
  {journal} {Phys. Rev. Lett.}\ }\textbf {\bibinfo {volume} {128}},\ \bibinfo
  {pages} {042301} (\bibinfo {year} {2022})},\ \Eprint
  {http://arxiv.org/abs/2111.02908} {arXiv:2111.02908 [nucl-th]} \BibitemShut
  {NoStop}%
\bibitem [{\citenamefont {Acharya}\ \emph {et~al.}(2022)\citenamefont {Acharya}
  \emph {et~al.}}]{ALICE:2021gxt}%
  \BibitemOpen
  \bibfield  {author} {\bibinfo {author} {\bibfnamefont {S.}~\bibnamefont
  {Acharya}} \emph {et~al.} (\bibinfo {collaboration} {ALICE}),\ }\href
  {\doibase 10.1016/j.physletb.2022.137393} {\bibfield  {journal} {\bibinfo
  {journal} {Phys. Lett. B}\ }\textbf {\bibinfo {volume} {834}},\ \bibinfo
  {pages} {137393} (\bibinfo {year} {2022})},\ \Eprint
  {http://arxiv.org/abs/2111.06106} {arXiv:2111.06106 [nucl-ex]} \BibitemShut
  {NoStop}%
\bibitem [{\citenamefont {Nijs}\ and\ \citenamefont {van~der
  Schee}(2022)}]{Nijs:2022rme}%
  \BibitemOpen
  \bibfield  {author} {\bibinfo {author} {\bibfnamefont {G.}~\bibnamefont
  {Nijs}}\ and\ \bibinfo {author} {\bibfnamefont {W.}~\bibnamefont {van~der
  Schee}},\ }\href {\doibase 10.1103/PhysRevLett.129.232301} {\bibfield
  {journal} {\bibinfo  {journal} {Phys. Rev. Lett.}\ }\textbf {\bibinfo
  {volume} {129}},\ \bibinfo {pages} {232301} (\bibinfo {year} {2022})},\
  \Eprint {http://arxiv.org/abs/2206.13522} {arXiv:2206.13522 [nucl-th]}
  \BibitemShut {NoStop}%
\bibitem [{\citenamefont {M\"antysaari}\ \emph {et~al.}(2017)\citenamefont
  {M\"antysaari}, \citenamefont {Schenke}, \citenamefont {Shen},\ and\
  \citenamefont {Tribedy}}]{Mantysaari:2017cni}%
  \BibitemOpen
  \bibfield  {author} {\bibinfo {author} {\bibfnamefont {H.}~\bibnamefont
  {M\"antysaari}}, \bibinfo {author} {\bibfnamefont {B.}~\bibnamefont
  {Schenke}}, \bibinfo {author} {\bibfnamefont {C.}~\bibnamefont {Shen}}, \
  and\ \bibinfo {author} {\bibfnamefont {P.}~\bibnamefont {Tribedy}},\ }\href
  {\doibase 10.1016/j.physletb.2017.07.038} {\bibfield  {journal} {\bibinfo
  {journal} {Phys. Lett. B}\ }\textbf {\bibinfo {volume} {772}},\ \bibinfo
  {pages} {681} (\bibinfo {year} {2017})},\ \Eprint
  {http://arxiv.org/abs/1705.03177} {arXiv:1705.03177 [nucl-th]} \BibitemShut
  {NoStop}%
\bibitem [{\citenamefont {Welsh}\ \emph {et~al.}(2016)\citenamefont {Welsh},
  \citenamefont {Singer},\ and\ \citenamefont {Heinz}}]{Welsh:2016siu}%
  \BibitemOpen
  \bibfield  {author} {\bibinfo {author} {\bibfnamefont {K.}~\bibnamefont
  {Welsh}}, \bibinfo {author} {\bibfnamefont {J.}~\bibnamefont {Singer}}, \
  and\ \bibinfo {author} {\bibfnamefont {U.~W.}\ \bibnamefont {Heinz}},\ }\href
  {\doibase 10.1103/PhysRevC.94.024919} {\bibfield  {journal} {\bibinfo
  {journal} {Phys. Rev. C}\ }\textbf {\bibinfo {volume} {94}},\ \bibinfo
  {pages} {024919} (\bibinfo {year} {2016})},\ \Eprint
  {http://arxiv.org/abs/1605.09418} {arXiv:1605.09418 [nucl-th]} \BibitemShut
  {NoStop}%
\bibitem [{\citenamefont {Schenke}\ and\ \citenamefont
  {Venugopalan}(2014)}]{Schenke:2014zha}%
  \BibitemOpen
  \bibfield  {author} {\bibinfo {author} {\bibfnamefont {B.}~\bibnamefont
  {Schenke}}\ and\ \bibinfo {author} {\bibfnamefont {R.}~\bibnamefont
  {Venugopalan}},\ }\href {\doibase 10.1103/PhysRevLett.113.102301} {\bibfield
  {journal} {\bibinfo  {journal} {Phys. Rev. Lett.}\ }\textbf {\bibinfo
  {volume} {113}},\ \bibinfo {pages} {102301} (\bibinfo {year} {2014})},\
  \Eprint {http://arxiv.org/abs/1405.3605} {arXiv:1405.3605 [nucl-th]}
  \BibitemShut {NoStop}%
\bibitem [{\citenamefont {Zheng}\ \emph {et~al.}(2021)\citenamefont {Zheng},
  \citenamefont {Zhang}, \citenamefont {Liu}, \citenamefont {Lin},
  \citenamefont {Shou},\ and\ \citenamefont {Yin}}]{Zheng:2021jrr}%
  \BibitemOpen
  \bibfield  {author} {\bibinfo {author} {\bibfnamefont {L.}~\bibnamefont
  {Zheng}}, \bibinfo {author} {\bibfnamefont {G.-H.}\ \bibnamefont {Zhang}},
  \bibinfo {author} {\bibfnamefont {Y.-F.}\ \bibnamefont {Liu}}, \bibinfo
  {author} {\bibfnamefont {Z.-W.}\ \bibnamefont {Lin}}, \bibinfo {author}
  {\bibfnamefont {Q.-Y.}\ \bibnamefont {Shou}}, \ and\ \bibinfo {author}
  {\bibfnamefont {Z.-B.}\ \bibnamefont {Yin}},\ }\href {\doibase
  10.1140/epjc/s10052-021-09527-5} {\bibfield  {journal} {\bibinfo  {journal}
  {Eur. Phys. J. C}\ }\textbf {\bibinfo {volume} {81}},\ \bibinfo {pages} {755}
  (\bibinfo {year} {2021})},\ \Eprint {http://arxiv.org/abs/2104.05998}
  {arXiv:2104.05998 [hep-ph]} \BibitemShut {NoStop}%
\bibitem [{\citenamefont {Zhao}\ \emph {et~al.}(2023)\citenamefont {Zhao},
  \citenamefont {Lin}, \citenamefont {Zheng},\ and\ \citenamefont
  {Ma}}]{Zhao:2021bef}%
  \BibitemOpen
  \bibfield  {author} {\bibinfo {author} {\bibfnamefont {X.-L.}\ \bibnamefont
  {Zhao}}, \bibinfo {author} {\bibfnamefont {Z.-W.}\ \bibnamefont {Lin}},
  \bibinfo {author} {\bibfnamefont {L.}~\bibnamefont {Zheng}}, \ and\ \bibinfo
  {author} {\bibfnamefont {G.-L.}\ \bibnamefont {Ma}},\ }\href {\doibase
  10.1016/j.physletb.2023.137799} {\bibfield  {journal} {\bibinfo  {journal}
  {Phys. Lett. B}\ }\textbf {\bibinfo {volume} {839}},\ \bibinfo {pages}
  {137799} (\bibinfo {year} {2023})},\ \Eprint
  {http://arxiv.org/abs/2112.01232} {arXiv:2112.01232 [nucl-th]} \BibitemShut
  {NoStop}%
\bibitem [{\citenamefont {Giacalone}()}]{Giacalone:2022hnz}%
  \BibitemOpen
  \bibfield  {author} {\bibinfo {author} {\bibfnamefont {G.}~\bibnamefont
  {Giacalone}},\ }\href@noop {} {\ }\Eprint {http://arxiv.org/abs/2208.06839}
  {arXiv:2208.06839 [nucl-th]} \BibitemShut {NoStop}%
\bibitem [{\citenamefont {Wang}\ and\ \citenamefont
  {Gyulassy}(1991)}]{Wang:1991hta}%
  \BibitemOpen
  \bibfield  {author} {\bibinfo {author} {\bibfnamefont {X.-N.}\ \bibnamefont
  {Wang}}\ and\ \bibinfo {author} {\bibfnamefont {M.}~\bibnamefont
  {Gyulassy}},\ }\href {\doibase 10.1103/PhysRevD.44.3501} {\bibfield
  {journal} {\bibinfo  {journal} {Phys. Rev. D}\ }\textbf {\bibinfo {volume}
  {44}},\ \bibinfo {pages} {3501} (\bibinfo {year} {1991})}\BibitemShut
  {NoStop}%
\bibitem [{\citenamefont {Zhang}(1998)}]{Zhang:1997ej}%
  \BibitemOpen
  \bibfield  {author} {\bibinfo {author} {\bibfnamefont {B.}~\bibnamefont
  {Zhang}},\ }\href {\doibase 10.1016/S0010-4655(98)00010-1} {\bibfield
  {journal} {\bibinfo  {journal} {Comput. Phys. Commun.}\ }\textbf {\bibinfo
  {volume} {109}},\ \bibinfo {pages} {193} (\bibinfo {year} {1998})},\ \Eprint
  {http://arxiv.org/abs/nucl-th/9709009} {arXiv:nucl-th/9709009} \BibitemShut
  {NoStop}%
\bibitem [{\citenamefont {Li}\ and\ \citenamefont {Ko}(1995)}]{Li:1995pra}%
  \BibitemOpen
  \bibfield  {author} {\bibinfo {author} {\bibfnamefont {B.-A.}\ \bibnamefont
  {Li}}\ and\ \bibinfo {author} {\bibfnamefont {C.~M.}\ \bibnamefont {Ko}},\
  }\href {\doibase 10.1103/PhysRevC.52.2037} {\bibfield  {journal} {\bibinfo
  {journal} {Phys. Rev. C}\ }\textbf {\bibinfo {volume} {52}},\ \bibinfo
  {pages} {2037} (\bibinfo {year} {1995})},\ \Eprint
  {http://arxiv.org/abs/nucl-th/9505016} {arXiv:nucl-th/9505016} \BibitemShut
  {NoStop}%
\bibitem [{\citenamefont {Liu}\ \emph {et~al.}(2024)\citenamefont {Liu},
  \citenamefont {Li}, \citenamefont {Wang}, \citenamefont {Xu}, \citenamefont
  {Ren},\ and\ \citenamefont {Huang}}]{Liu:2023gun}%
  \BibitemOpen
  \bibfield  {author} {\bibinfo {author} {\bibfnamefont {L.-M.}\ \bibnamefont
  {Liu}}, \bibinfo {author} {\bibfnamefont {S.-J.}\ \bibnamefont {Li}},
  \bibinfo {author} {\bibfnamefont {Z.}~\bibnamefont {Wang}}, \bibinfo {author}
  {\bibfnamefont {J.}~\bibnamefont {Xu}}, \bibinfo {author} {\bibfnamefont
  {Z.-Z.}\ \bibnamefont {Ren}}, \ and\ \bibinfo {author} {\bibfnamefont
  {X.-G.}\ \bibnamefont {Huang}},\ }\href {\doibase
  10.1016/j.physletb.2024.138724} {\bibfield  {journal} {\bibinfo  {journal}
  {Phys. Lett. B}\ }\textbf {\bibinfo {volume} {854}},\ \bibinfo {pages}
  {138724} (\bibinfo {year} {2024})},\ \Eprint
  {http://arxiv.org/abs/2312.13572} {arXiv:2312.13572 [nucl-th]} \BibitemShut
  {NoStop}%
\bibitem [{\citenamefont {Wang}\ \emph
  {et~al.}(2024{\natexlab{b}})\citenamefont {Wang}, \citenamefont {Li},
  \citenamefont {Liu}, \citenamefont {Xu},\ and\ \citenamefont
  {Ren}}]{Wang:2024ulq}%
  \BibitemOpen
  \bibfield  {author} {\bibinfo {author} {\bibfnamefont {H.-C.}\ \bibnamefont
  {Wang}}, \bibinfo {author} {\bibfnamefont {S.-J.}\ \bibnamefont {Li}},
  \bibinfo {author} {\bibfnamefont {L.-M.}\ \bibnamefont {Liu}}, \bibinfo
  {author} {\bibfnamefont {J.}~\bibnamefont {Xu}}, \ and\ \bibinfo {author}
  {\bibfnamefont {Z.-Z.}\ \bibnamefont {Ren}},\ }\href {\doibase
  10.1103/PhysRevC.110.034909} {\bibfield  {journal} {\bibinfo  {journal}
  {Phys. Rev. C}\ }\textbf {\bibinfo {volume} {110}},\ \bibinfo {pages}
  {034909} (\bibinfo {year} {2024}{\natexlab{b}})},\ \Eprint
  {http://arxiv.org/abs/2409.02452} {arXiv:2409.02452 [nucl-th]} \BibitemShut
  {NoStop}%
\bibitem [{\citenamefont {Xu}\ and\ \citenamefont {Ko}(2011)}]{Xu:2011fe}%
  \BibitemOpen
  \bibfield  {author} {\bibinfo {author} {\bibfnamefont {J.}~\bibnamefont
  {Xu}}\ and\ \bibinfo {author} {\bibfnamefont {C.~M.}\ \bibnamefont {Ko}},\
  }\href {\doibase 10.1103/PhysRevC.84.014903} {\bibfield  {journal} {\bibinfo
  {journal} {Phys. Rev. C}\ }\textbf {\bibinfo {volume} {84}},\ \bibinfo
  {pages} {014903} (\bibinfo {year} {2011})},\ \Eprint
  {http://arxiv.org/abs/1103.5187} {arXiv:1103.5187 [nucl-th]} \BibitemShut
  {NoStop}%
\bibitem [{\citenamefont {Lim}\ \emph {et~al.}(2019)\citenamefont {Lim},
  \citenamefont {Hu}, \citenamefont {Belmont}, \citenamefont {Hill},
  \citenamefont {Nagle},\ and\ \citenamefont
  {Perepelitsa}}]{PhysRevC.100.024908}%
  \BibitemOpen
  \bibfield  {author} {\bibinfo {author} {\bibfnamefont {S.~H.}\ \bibnamefont
  {Lim}}, \bibinfo {author} {\bibfnamefont {Q.}~\bibnamefont {Hu}}, \bibinfo
  {author} {\bibfnamefont {R.}~\bibnamefont {Belmont}}, \bibinfo {author}
  {\bibfnamefont {K.~K.}\ \bibnamefont {Hill}}, \bibinfo {author}
  {\bibfnamefont {J.~L.}\ \bibnamefont {Nagle}}, \ and\ \bibinfo {author}
  {\bibfnamefont {D.~V.}\ \bibnamefont {Perepelitsa}},\ }\href {\doibase
  10.1103/PhysRevC.100.024908} {\bibfield  {journal} {\bibinfo  {journal}
  {Phys. Rev. C}\ }\textbf {\bibinfo {volume} {100}},\ \bibinfo {pages}
  {024908} (\bibinfo {year} {2019})}\BibitemShut {NoStop}%
\bibitem [{\citenamefont {Abdulameer}\ \emph {et~al.}(2023)\citenamefont
  {Abdulameer} \emph {et~al.}}]{PhysRevC.107.024907}%
  \BibitemOpen
  \bibfield  {author} {\bibinfo {author} {\bibfnamefont {N.~J.}\ \bibnamefont
  {Abdulameer}} \emph {et~al.} (\bibinfo {collaboration} {PHENIX
  Collaboration}),\ }\href {\doibase 10.1103/PhysRevC.107.024907} {\bibfield
  {journal} {\bibinfo  {journal} {Phys. Rev. C}\ }\textbf {\bibinfo {volume}
  {107}},\ \bibinfo {pages} {024907} (\bibinfo {year} {2023})}\BibitemShut
  {NoStop}%
\bibitem [{\citenamefont {Abdulhamid}\ \emph {et~al.}(2023)\citenamefont
  {Abdulhamid} \emph {et~al.}}]{PhysRevLett.130.242301}%
  \BibitemOpen
  \bibfield  {author} {\bibinfo {author} {\bibfnamefont {M.~I.}\ \bibnamefont
  {Abdulhamid}} \emph {et~al.} (\bibinfo {collaboration} {STAR
  Collaboration}),\ }\href {\doibase 10.1103/PhysRevLett.130.242301} {\bibfield
   {journal} {\bibinfo  {journal} {Phys. Rev. Lett.}\ }\textbf {\bibinfo
  {volume} {130}},\ \bibinfo {pages} {242301} (\bibinfo {year}
  {2023})}\BibitemShut {NoStop}%
\bibitem [{\citenamefont {Bozek}(2016)}]{Bozek:2016yoj}%
  \BibitemOpen
  \bibfield  {author} {\bibinfo {author} {\bibfnamefont {P.}~\bibnamefont
  {Bozek}},\ }\href {\doibase 10.1103/PhysRevC.93.044908} {\bibfield  {journal}
  {\bibinfo  {journal} {Phys. Rev. C}\ }\textbf {\bibinfo {volume} {93}},\
  \bibinfo {pages} {044908} (\bibinfo {year} {2016})},\ \Eprint
  {http://arxiv.org/abs/1601.04513} {arXiv:1601.04513 [nucl-th]} \BibitemShut
  {NoStop}%
\bibitem [{\citenamefont {Jia}\ \emph {et~al.}(2022)\citenamefont {Jia},
  \citenamefont {Huang},\ and\ \citenamefont {Zhang}}]{PhysRevC.105.014906}%
  \BibitemOpen
  \bibfield  {author} {\bibinfo {author} {\bibfnamefont {J.}~\bibnamefont
  {Jia}}, \bibinfo {author} {\bibfnamefont {S.}~\bibnamefont {Huang}}, \ and\
  \bibinfo {author} {\bibfnamefont {C.}~\bibnamefont {Zhang}},\ }\href
  {\doibase 10.1103/PhysRevC.105.014906} {\bibfield  {journal} {\bibinfo
  {journal} {Phys. Rev. C}\ }\textbf {\bibinfo {volume} {105}},\ \bibinfo
  {pages} {014906} (\bibinfo {year} {2022})}\BibitemShut {NoStop}%
\bibitem [{\citenamefont {M\"antysaari}\ \emph {et~al.}(2024)\citenamefont
  {M\"antysaari}, \citenamefont {Schenke}, \citenamefont {Shen},\ and\
  \citenamefont {Zhao}}]{Mantysaari:2024uwn}%
  \BibitemOpen
  \bibfield  {author} {\bibinfo {author} {\bibfnamefont {H.}~\bibnamefont
  {M\"antysaari}}, \bibinfo {author} {\bibfnamefont {B.}~\bibnamefont
  {Schenke}}, \bibinfo {author} {\bibfnamefont {C.}~\bibnamefont {Shen}}, \
  and\ \bibinfo {author} {\bibfnamefont {W.}~\bibnamefont {Zhao}},\ }\href
  {\doibase 10.1103/PhysRevC.110.054913} {\bibfield  {journal} {\bibinfo
  {journal} {Phys. Rev. C}\ }\textbf {\bibinfo {volume} {110}},\ \bibinfo
  {pages} {054913} (\bibinfo {year} {2024})},\ \Eprint
  {http://arxiv.org/abs/2409.19064} {arXiv:2409.19064 [nucl-th]} \BibitemShut
  {NoStop}%
\bibitem [{\citenamefont {Abdallah}\ \emph {et~al.}(2022)\citenamefont
  {Abdallah} \emph {et~al.}}]{STAR:2021mii}%
  \BibitemOpen
  \bibfield  {author} {\bibinfo {author} {\bibfnamefont {M.}~\bibnamefont
  {Abdallah}} \emph {et~al.} (\bibinfo {collaboration} {STAR}),\ }\href
  {\doibase 10.1103/PhysRevC.105.014901} {\bibfield  {journal} {\bibinfo
  {journal} {Phys. Rev. C}\ }\textbf {\bibinfo {volume} {105}},\ \bibinfo
  {pages} {014901} (\bibinfo {year} {2022})},\ \Eprint
  {http://arxiv.org/abs/2109.00131} {arXiv:2109.00131 [nucl-ex]} \BibitemShut
  {NoStop}%
\end{thebibliography}%
\end{document}